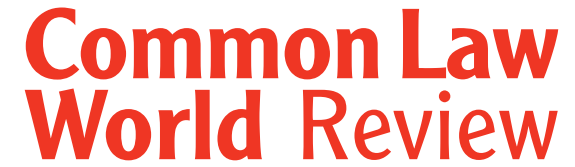

# Unjust enrichment as a remedy for AI's unauthorised use of protected data

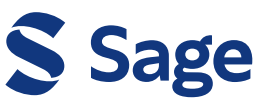

**Yangzi Li[1] and Jyh-An Lee[2]**

**Abstract**

The unauthorised use of data in the training of generative AI models presents significant legal challenges, particularly under intellectual property (IP) and privacy laws. These frameworks frequently grapple with the intricate relationship between data ownership and AI innovation, resulting in ongoing debates regarding optimal protection and enforceability. This article delves into the considerable potential of unjust enrichment as an alternative legal doctrine for resolving disputes arising from such unauthorised data use. We explore how the concept of unjust enrichment captures the wrongfulness of unauthorised data use in a manner distinct from IP infringement and privacy violations. Furthermore, we analyse the extent to which gain-based restitution for unjust enrichment may prove more advantageous than existing remedies, including legal, equitable and statutory options. We contend that by shifting the emphasis from establishing wrongful conduct to recovering benefits obtained unjustly, unjust enrichment offers a pragmatic and equitable framework that reconciles the rights of data owners with the interests of AI developers.

## Introduction

Data is the lifeblood of generative AI (GenAI) models, including large language models (LLMs), diffusion models and biometric recognition systems. LLMs like GPT and

[1]Centre for Technology, Robotics, Artificial Intelligence & the Law, Faculty of Law, National University of Singapore, Singapore, Singapore
[2]Faculty of Law, The Chinese University of Hong Kong, Hong Kong, Hong Kong

**Corresponding author:**
Jyh-An Lee, Faculty of Law, The Chinese University of Hong Kong, Hong Kong, Hong Kong.
Email: jalee@cuhk.edu.hk

BERT are trained on diverse datasets sourced from websites, books and various digital sources.[1] Diffusion models, such as DALL-E 3 and Stable Diffusion, rely on large-scale datasets of image–text pairs obtained from the internet.[2] Biometric recognition models like Clearview AI and NEC Fingerprint Recognition are typically trained using biometric datasets collected from government databases, social media platforms and user-contributed data.[3] It is important to note that these training datasets often contain protected information, such as copyright works and personal data, raising significant legal concerns regarding the unauthorised use of such data and potential conflicts between data owners and AI developers.[4]

The current disputes primarily centre around data owner's copyright, publicity right[5] and privacy right.[6] Ongoing AI-related lawsuits in the United States (US) showcase data owners' efforts to seek remedies, including damages and injunctions, for the unauthorised use of their data in AI training. Some cases have seen judges dismissing claims of copyright infringement, violation of publicity rights and invasion of privacy due to the lack of substantial evidence presented by the plaintiffs. For instance, in *Andersen v. Stability AI*, Judge Orrick dismissed some of the plaintiffs' direct copyright infringement allegations and the right of publicity claims because of their failure to clarify relevant doctrines and fact in support those allegations.[7] Further, in *Doe 1 v. GitHub*, Judge Tigar ruled that the plaintiffs failed to demonstrate a sufficient injury in fact to establish standing for privacy violations and were unable to seek damages due to the lack of specific injury.[8] This trend points to the challenges and uncertainties in adequately compensating data owners through existing intellectual property (IP) and privacy regimes. Moreover, pending lawsuits indicate courts' cautious examination of the legality of unauthorised use of various data by AI developers. Courts may be hesitant to label such uses as wrongful, as doing so could potentially impede the progress of AI technologies. The dilemma lies in finding a balance between providing adequate remedies for claimants without overly hindering AI innovation.

While most current lawsuits have been filed in the US, the underlying legal and policy issues transcend national boundaries. The tension between protecting the rights of data owners and incentivising AI innovation has existed across various jurisdictions. This article explores how common law might address these challenges. In response to these challenges, scholars are exploring the potential of restitution for unjust enrichment as a remedy for the unauthorised use of protected data.[9] Unjust enrichment, in general, arises when one party benefits unfairly at the expense of another, deemed

[1] Tom B. Brown, 'Language Models Are Few-Shot Learners' (2020) arXiv preprint arXiv:2005.14165.

[2] Reed Scott and others, 'Generative Adversarial Text to Image Synthesis' (2016) International Conference on Machine Learning 1060, 1060–69, PMLR.

[3] See, for example, Robert Hart, ''Clearview AI – Controversial Facial Recognition Firm – Fined $33 Million For 'Illegal Database' (*Forbes*, 3 September 2024), <https://www.forbes.com/sites/roberthart/2024/09/03/clearview-ai-controversial-facial-recognition-firm-fined-33-million-for-illegal-database/>; Joel R. McConvey, 'NEC Updates NeoFace Reveal with New Interface and Facial Recognition Tools' *(Biometric Update*, 19 March 2024), <https://www.biometricupdate.com/202403/nec-updates-neoface-reveal-with-new-interface-and-facial-recognition-tools>

[4] Amanda Heidt, 'Intellectual Property and Data Privacy: The Hidden Risks of AI' (*Nature,* 4 September 2024), <https://www.nature.com/articles/d41586-024-02838-z>.

[5] See, for example, *Andersen et al. v. Stability AI, Inc.*, No. 3:23-cv-00201 (N.D. Cal. filed Jan. 13, 2023).

[6] See, for example, *Doe 1 v. GitHub, Inc.*, 672 F. Supp. 3d 837, 849–50 (N.D. Cal. 2023) (Judge Tigar grants leave to amend to the claims of violation of privacy law as the plaintiff failed to provide insufficient standing).

[7] *Andersen et al. v. Stability AI, Inc.*, No. 3:23-cv-00201-WHO, 9–10, 13–14, 19–20 (N.D. Cal. filed Oct. 30, 2023).

[8] *Doe 1 v. GitHub, Inc.*, 672 F. Supp. 3d 837, 840–45.

[9] See, for example, Ying Hu, 'Mainstreaming Unjust Enrichment and Restitution in Data Security Law' (2023) 13 UC Irvine L Rev 855; Frank Pasquale and Haochen Sun, 'Consent and Compensation: Resolving Generative AI's Copyright Crisis' (2024) 110 Virginia L Rev Online 207; Benjamin L. W. Sobel, 'A New Common Law of Web Scraping' (2021)

inequitable in nature.[10] It is considered 'quasi-contractual,' distinct from both tort and contract within the law of obligations.[11] Unjust enrichment stands apart from both wrongdoing and contractual obligations.[12] A right to restitution thereby arises from unjust enrichment, enabling the award of restitution to the aggrieved party.[13] In this paper, we argue that applying this principle to the unauthorised use of data to train AI models could not only complement existing IP and privacy laws but also foster a balance between the rights of data owners and the imperative for AI innovation.[14] However, current literature has not fully explored whether unjust enrichment could effectively capture the wrongfulness and supplement the often inadequate remedies in IP and privacy laws for the unauthorised use of protected data in GenAI training. This article delves into the potential for unjust enrichment to address the problem of unauthorised data usage in training GenAI models. Part 2 examines claims of unjust enrichment against AI developers through existing lawsuits. Part 3 evaluates whether the unauthorised use of protected data by AI developers constitutes unjust enrichment. Part 4 discusses how the law of unjust enrichment and its attached restitution could provide an alternative solution other than doctrines and remedies under current IP law and privacy law. Part 5 concludes the article.

## Unjust enrichment claims against AI developers

While unjust enrichment claims often appear alongside a series of other causes of action,[15] whether unjust enrichment is a standalone cause of action remains inconsistent and unclear in the US case law.[16] For example, in *Hartford v. J.R. Mktg*, the California Supreme Court held that unjust enrichment is an independent cause of action separate from a breach of contract claim.[17] Conversely, the Ninth Circuit, in *Astiana v. Hain Celestial*, characterised unjust enrichment as 'a quasi-contract claim for restitution' rather than a distinct cause of action.[18] In contrast, commonwealth jurisdictions generally recognise unjust enrichment as an independent and residual cause of action.[19] More specifically, the claim is typically invoked in circumstances that

25 Lewis & Clark L Rev 147; Yotam Kaplan and Gordon-Tapiero, Ayelet, 'Generative AI Training as Unjust Enrichment'(2025) 86 Ohio State L J (forthcoming).

[10] The definitions and frameworks developed by Goff and Jones, Peter Birks and Andrew Burrows have had a profound influence on how modern UK courts approach unjust enrichment. While their definitions differ on certain points, this article adopts the most widely accepted formulations within the common law tradition. See, for example, Charles Mitchell, Paul Mitchell and Stephen Watterson(eds), *Goff & Jones: The Law of Unjust Enrichment* (10th edn, Sweet & Maxwell 2022); Andrew Burrows, *A Restatement of the English Law of Unjust Enrichment* (OUP 2012); Peter Birks, *Unjust Enrichment* (2nd edn, OUP 2005).

[11] *Curtin v. Star Ed. Inc.*, 2 F. Supp. 2d 670, 674 (E.D. Pa. 1998); *Hershey Foods Corp. v. Ralph Chapek, Inc.*, 828 F.2d 989, 999 (3d Cir.1987).

[12] Peter Birks, 'Unjust Enrichment and Wrongful Enrichment' (2001) 79 Texas L Rev1767, 1794.

[13] ibid 4.

[14] Gordon-Tapiero and Kaplan (n 9), 4–5.

[15] Bernard Chao, 'Privacy Losses as Wrongful Gains' (2021) 106 Iowa L Rev 555, 572.

[16] *Bruton v. Gerber Products Company*, 703 F. Appx. 468, 470 (9th Cir. 2017) ('At the time when the district court dismissed this claim, California's caselaw on whether unjust enrichment could be sustained as a standalone cause of action was uncertain and inconsistent. But since then, the California Supreme Court has clarified California law, allowing an independent claim for unjust enrichment to proceed in an insurance dispute'.).

[17] *Hartford Cas. Ins. Co. v. J.R. Mktg.*, LLC, 353 P.3d 319, 326 (Cal. 2015).

[18] *Astiana v. Hain Celestial Grp., Inc.*, 783 F.3d 753, 762 (9th Cir. 2015).

[19] Peter Birks, 'The Law of Unjust Enrichment: A Millennial Resolution' (1999) Singapore JLS 318, 328; Charlie Webb, 'What Is Unjust Enrichment?' (2009) 29(2) Oxford Journal of Legal Studies 215, 218.

warrant restitution solely on the basis that the enrichment is unjust – rather than arising from a tort, contract or any other legal basis.[20]

As GenAI models increasingly rely on diverse datasets, data owners are filing lawsuits for copyright infringement, right of publicity violation and privacy breaches against AI developers. In some of these cases, unjust enrichment is cited as one of the causes of action.[21] In particular, unjust enrichment is frequently coupled with allegations, such as copyright infringement,[22] the right of publicity violations[23] and the invasion of privacy.[24] This section examines the role of unjust enrichment in these lawsuits and highlights its potential in addressing the tension between data owners and AI developers.

### *Copyright*

Copyright owners frequently claim infringement against AI developers for using their copyright works to train GenAI models without a licence. Scholars and policymakers have explored possible solutions for such unauthorised uses within the copyright framework. Some are examining whether rules on copyright limitations and exceptions, such as the fair use doctrine in the US and text and data mining (TDM) exceptions in the European Union (EU) and United Kingdom (UK), could shield AI developers from infringement liability.[25] Others have proposed broadening the scope of compulsory licencing to ensure fair compensation for copyright holders while simultaneously fostering AI innovation.[26] Despite these efforts, there is no international consensus on the optimal approach to copyright law in the context of AI.

As an alternative approach, unjust enrichment, has emerged as a potential avenue for copyright owners seeking remedies. In several cases, unjust enrichment claims are raised in conjunction with allegations of copyright infringement and violations of the Digital Millennium Copyright Act (DMCA). Notably, claimants in these cases contend that unjust enrichment occurs during the AI training process, which requires the use of large quantities of copyright works. For example, in *Tremblay v. OpenAI*, the plaintiffs accused OpenAI of unjust enrichment by using their copyright works – on which they had expended substantial time and effort – to train AI models without permission.[27] This purportedly deprived the plaintiffs of the benefits of

[20] ibid.

[21] See, for example, *Chabon v. Meta Platforms, Inc.*, No. 3:23-cv-04663 (N.D. Cal. filed Sept. 12, 2023); *A.T. et al. v. OpenAI LP et al.*, No. 3:23-cv-04557 (N.D. Cal. filed Sept. 6, 2023); *Paul Tremblay et al. v. OpenAI, Inc. et al.*, No. 3:23-cv-03223 (N.D. Cal. filed July 7, 2023); *Kadrey v. Meta, Silverman v. OpenAI, Inc.*, No. 3:23-CV-03416 (N.D. Cal. filed July 7, 2023); *Andersen et al. v. Stability AI*, Inc., No. 3:23-cv-00201 (N.D. Cal. filed Jan. 13, 2023); *Doe 1 v. GitHub, Inc.*, No. 3:22-cv-06823 (N.D. Cal. filed Nov. 3, 2022).

[22] See, for example, *Chabon v. Meta Platforms, Inc.*, No. 3:23-cv-04663 (N.D. Cal. filed Sept. 12, 2023); *Andersen et al. v. Stability AI, Inc.*, No. 3:23-cv-00201 (N.D. Cal. filed Jan. 13, 2023).

[23] *Andersen et al. v. Stability AI, Inc.*, No. 3:23-cv-00201 (N.D. Cal. filed Jan. 13, 2023).

[24] See, for example, *A.T. et al. v. OpenAI LP et al.*, No. 3:23-cv-04557 (N.D. Cal. filed Sept. 6, 2023).

[25] See, for example, Matthew Sag and Peter K. Yu, 'The Globalization of Copyright Exceptions for AI Training' (2025) 74 Emory L J 1163; Tatsuhiro Ueno, 'The Flexible Copyright Exception for "Non Enjoyment" Purposes' (2021) 70 GRUR Int' L 145; Alina Trapova & João Pedro Quintais, 'The UK Government Moves Forward with a Text and Data Mining Exception for All Purposes,' *Kluwer Copyright Blog* (Aug. 24, 2022), <https://copyrightblog.kluweriplaw.com/2022/08/24/the-ukgovernment-moves-forward-with-a-text-and-data-mining-exception-for-all-purposes>.

[26] See, for example, Rita Matulionyte, 'Generative AI and Copyright: Exception, Compensation or Both?' (2023) 134 Intellectual Property Forum: Journal of the Intellectual and Industrial Property Society of Australia and New Zealand 1.

[27] *Paul Tremblay et al. v. OpenAI, Inc. et al.*, No. 3:23-cv-03223 (N.D. Cal. filed July 7, 2023).

their works, resulting in harm.[28] However, the US courts often rule that such claims are pre-empted by copyright law, as unauthorised use is already addressed through copyright infringement claims.[29] The notion that unjust enrichment claims are pre-empted by copyright infringement claims predates the advent of AI technologies.[30] Courts have held that attempts to seek damages for unauthorised copying or other activity covered by Section 106 are thinly veiled claims for damages.[31] In response to dismissals based on pre-emption, some copyright owners are pursuing unjust enrichment claims in new ways. In *Millette v. OpenAI*, the plaintiff did not claim copyright infringement against OpenAI but instead accused the company of only unjust enrichment. The plaintiff argued that OpenAI had been unjustly enriched by profiting from the sale of AI products trained on the plaintiff's works without providing any compensation.[32] In contrast to the *Tremblay* case, the plaintiff in *Millette* stressed that there was no adequate legal remedy available to address the defendant's unjust conduct.[33] The plaintiff further contended that restitutions should be separate from damages, and the court could order restitution even if there was insufficient evidence to support damages.[34]

It is worth noting that the US adopts a rather distinctive approach to the pre-emption of unjust enrichment by copyright law, a practice that is largely absent in most common law jurisdictions. For instance, English law allows recovery under unjust enrichment even in the presence of a copyright infringement claim, thereby providing greater flexibility in addressing such matters.[35]

### *The right of publicity*

The right of publicity pertains to an individual's authority over the commercial use of their identity, encompassing as aspects such as their name, voice, image, gestures and likeness as recognised under the US common law.[36] Originally grounded in privacy law, the right of publicity has evolved into a distinct realm within IP law.[37] An infringement of the right of publicity

[28] ibid.

[29] *Kadrey et al. v. Meta Platforms, Inc.*, No. 23-cv-03417-VC (N.D. Cal. filed July 2023), at 4 ('the duty alleged (to the extent it could be thought to exist) is premised on the rights protected by the Copyright Act and thus any claims for breach of this duty are preempted'.); *Doe 1 v. GitHub, Inc.*, 672 F. Supp. 3d 837 (N.D. Cal. 2023); *Del Madera Props. V. Rhodes & Gardner, Inc.*, 820 F.2d 973, 977 (9th Cir. 1987) (finding pre-emption because an 'implied promise not to use or copy materials within the subject matter of copyright is equivalent' to the Copyright Act's protections); *Firoozye v. Earthlink Network*, 153 F. Supp. 2d 1115, 1126 (N.D. Cal. 2001) (finding pre-emption where plaintiff alleged defendant improperly benefitted from using copyright software and 'that a contract should be implied in law (e.g., a quasi-contract or unjust enrichment claim)').

[30] Jennifer Geller, 'A Mixtape Dj's Drama: An Argument for Copyright Preemption of Georgia's Unauthorized Reproduction Law' (2008) 8 Chi.-Kent J Intell Prop 1, 36; David E. Shipley and Jeffrey S. Hay, 'Protecting Research: Copyright, Common-Law Alternatives, and Federal Preemption' (1984)63 NC L Rev 125, 177(noting that 'the elements for the state law cause of action are equivalent to federal copyright claims').

[31] See, for example, *Amador v. McDonald's Corp.*, 601 F.Supp.2d 403, 408–409 (D.P.R. 2009); *Briarpatch Ltd., L.P. v. Phoenix Pictures, Inc.*, 373 F.3d 296, 305–307 (2d Cir. 2004); *Martin v. Walt Disney Internet Grp.*, No. 09CV1601-MMA, 2010 WL 2634695, at 6 (S.D. Cal. June 30, 2010); 1 William F. Patry, Patry on Copyright: Preemption of Other Laws, §18:42; §18:51 (2007).

[32] *David Millette v. OpenAI*, No. 3:24-cv-04710-TSH (N.D. Cal. filed Aug. 12, 2024).

[33] ibid 9.

[34] ibid.

[35] Danie Visser, 'Unjust Enrichment: A Study of Private Law and Public Values' (1999)3(3) Edin LR 402, 403.

[36] 1 J. Thomas Mccarthy and Roger E. Schechter, *The Rights of Publicity and Privacy* § 1:3 (2d ed. 2018); Ind. Code Ann. § 32-36-1-7 (West 1994).

[37] Mccarthy and Schechter (n 36) § 1:7; Jennifer E. Rothman, *The Right of Publicity: Privacy Reimagined for a Public World* (Harvard Univ. Press 2018) 11; Michael Henry (ed.), *International Privacy, Publicity and Personality Laws*

can be established if the plaintiffs can demonstrate the appropriation of their identity for another's commercial purposes, the absence of consent and the resulting injury.[38] Both private individuals and public figures are entitled to various remedies, such as damages and injunctive relief, for the unauthorised commercial exploitation of their identities.[39]

In the context of GenAI, individuals have similarly contended that unauthorised use of protected data containing personal identity for training AI models infringes the right of publicity. In *Lehrman et al. v. Lovo Inc.*, the plaintiffs filed a lawsuit against the defendant for exploiting their voices and/or identities without a licence, employing these elements to train a text-to-speech AI model and subsequently marketing the misappropriated voices as part of its service offering.[40] Notably, the plaintiffs lodged a claim under the False Advertising Act in the US, citing the right of publicity rather than pursuing an independent claim based solely on the right of publicity. In their complaint, the plaintiffs also seek several remedies, including damages, injunctive relief and both declaratory and equitable relief, such as restitution.[41]

Another illustrative case is the AI-generated song 'Heart on My Sleeve', which mimicked the voices of Drake and The Weeknd, causing a stir in the music industry and prompting legal inquiries into the unauthorised use of celebrity voices for training GenAI models.[42] This incident also ignited ethical debates regarding the potential harm inflicted on both musicians and consumers by AI-generated vocals.[43] Another notable instance involves Scarlett Johansson, who expressed reservations about OpenAI's chatbot bearing resemblance to her voice.[44] The adequacy of current laws in safeguarding individuals' vocal likeness and public personas within the realm of GenAI remains uncertain.[45]

The rise of deepfakes, which involve using AI to create realistic images or videos of individuals,[46] has introduced new complexities in the realm of publicity rights various other legal issues concerning fraud, blackmail, impersonation and propaganda.[47] Current publicity laws

(Butterworths 2001) 476; Huw Beverly-Smith, *The Commercial Appropriation of Personality* (CUP 2002) 145; Andrew Beckerman-Rodau, 'Toward a Limited Right of Publicity: An Argument for the Convergence of the Right of Publicity, Unfair Competition and Trademark Law' (2012) 23 Fordham Intell Prop Media & Ent LJ 132, 132.

[38] *Dancel v. Groupon, Inc.*, 949 F.3d 999 (7th Cir. 2019); Restatement (Third) of Unfair Competition § 46 (1995).

[39] Mccarthy and Schechter (n 36) § 1:3 ('The right of publicity is not merely a legal right of the "celebrity," but is a right inherent to everyone to control the commercial use of identity and persona and recover in court damages and the commercial value of an unpermitted taking'.).

[40] *Lehrman et al v. Lovo Inc*, No 1:24-cv-03770 (SDNY, July, 10 2025).

[41] ibid.

[42] Ben Sisario, 'An A.I. Hit of Fake "Drake" and "The Weeknd" Rattles the Music World' (*The New York Times,* 20 April 2023) <https://www.nytimes.com/2023/04/20/arts/music/ai-drake-weeknd-song.html> accessed 16 September 2024; 'An A.I. Hit of Fake 'Drake' and 'The Weeknd' Rattles the Music World' Chloe Veltman, 'When you realize your favorite new song was written and performed by … AI' (npr, 21 April 2023), <https://www.npr.org/2023/04/21/1171032649/ai-music-heart-on-my-sleeve-drake-the-weeknd> accessed 16 September 2024.

[43] ibid.

[44] Nicola Jones, 'Who Owns Your Voice? Scarlett Johansson OpenAI Complaint Raises Questions' (*Nature*, 29 May 2024) <https://doi.org/10.1038/d41586-024-01578-4> accessed 16 September 2024.

[45] ibid.

[46] Sara H. Jodka, 'Manipulating Reality: The Intersection Of Deepfakes And The Law' (*Reuters,* 1 February 2024) <https://www.reuters.com/legal/legalindustry/manipulating-reality-intersection-deepfakes-law-2024-02-01/> accessed 16 September 2024.

[47] ibid; see also Quentin J. Ullrich, 'Is This Video Real? The Principal Mischief of Deepfakes and How the Lanham Act Can Address It,' (2021) 55 Colum JL & Soc Probs 1, 5.

primarily target commercial exploitation such as advertising, often leaving gaps in addressing the non-commercial misuse of deepfakes.[48] The requirement of consent from the depicted individual for generating deepfakes in non-commercial settings remains ambiguous.[49] For example, Meta AI has introduced AI-generated icons and influencers on social media platforms, resembling real celebrities in appearance and voice.[50] Consequently, the right of publicity is encountering challenges in adequately compensating the celebrities depicted in the AI-generated deepfakes.[51]

Given the limitations of current remedial options, right-of-publicity holders have also sought recovery through unjust enrichment, though the effectiveness of this approach remains uncertain. In *Lehrman et al. v. Lovo Inc.*, plaintiffs claimed unjust enrichment, alongside claims of unfair competition and civil rights violations, alleging the misappropriation of their voices without permission or compensation.[52] They sought remedies including damages and restitution.[53] However, Judge Oetken dismissed the claim of unjust enrichment because it is pre-empted by New York's right of publicity statute.[54] Such pre-emption highly restricts unjust enrichment as a cause of action.

In *Andersen v. Stability AI*, the plaintiffs raised allegations of right-of-publicity violations, asserting that their names and 'artistic identities' – specifically their art and distinctive styles – were misused for commercial purposes.[55] However, the court dismissed these claims with the opportunity for amendment, citing the plaintiffs' failure to adequately expound on their right-of-publicity arguments or present plausible details regarding whether each defendant used their names for advertising or other commercial purposes.[56] Notably, the plaintiffs, among other claims, accused the defendant of infringing upon the right of publicity and requested damages to compensate both themselves and the class for harms resulting from the defendants' unjust enrichment.[57] Nevertheless, the claims did not establish a clear connection between unjust enrichment and the purported breach of the right of publicity, leaving this connection unexplained.

It is important to note that English law does not recognise a general right of publicity or personality.[58] For celebrities in common law jurisdictions like the UK, they often rely on the action of passing off to control the unauthorised use of their identities.[59]

[48] ibid.
[49] Alice Preminger and Matthew B. Kugler, 'The Right of Publicity Can Save Actors from Deepfake Armageddon' (2024) 39(2) Berkeley Tech L J 784, 803.
[50] 'Introducing New AI Experiences Across Our Family of Apps and Devices,' (*META,* 27 September 2023). <https://about.fb.com/news/2023/09/introducing-ai-powered-assistants-characters-and-creative-tools>.
[51] Reid M. Koski, 'Warhol, Drake, and Deepfakes: Monetizing the Right of Publicity in the Generative AI Era' (2024) 40 Ga St U L Rev 981, 985.
[52] *Lehrman et al v. Lovo Inc*, No 1:24-cv-03770 (SDNY, 10 July 2025).
[53] ibid.
[54] ibid 58.
[55] *Sarah Andersen v. Stability AI Inc* Case 3:23-cv-00201-WHO (ND Cal 2023), at 18.
[56] ibid 20.
[57] ibid, class action complaint, at 38, 43.
[58] See *Fenty v. Arcadia Group Brands Ltd* [2005] EWCA Civ 3, [29] (Kitchin LJ) ('There is in English law no "image right" or "character right" which allows a celebrity to control the use of his or her name or image'.).
[59] David Tan, 'Cultural Studies and the Common Law Passing Off Action' in *The Commercial Appropriation of Fame: A Cultural Analysis of the Right of Publicity and Passing Off* (Cambridge University Press 2017)199–245.

## *Privacy*

Data subject also claims violations of privacy rights against AI developers. GenAI models are increasing under scrutiny for privacy and data protection violations arising from their extensive use of personal data throughout the training process, including the memorisation of such data, user input prompts and the generated outputs.[60]

During the input phase, AI developers commonly employ web scraping to gather publicly available data containing personal information to train their models. Users of these models may also provide prompts, such as text or images, that include personal details to guide the AI's output generation process. For example, tech giant Meta notified a segment of its user base in the US, the UK and Europe that it would scrape public content shared on its social media platforms – such as Facebook, Instagram and WhatsApp – to train its GenAI models.[61] Given the diverse privacy and data protection regulations across different regions, Meta must obtain consent in specific jurisdictions for its data-scraping practices.[62] Additionally, users in countries with stricter privacy laws, such as the UK and European Union, generally have more opportunities to opt out of Meta's data-scraping activities.[63] Nevertheless, even in these regions, existing laws often fall short in providing adequate remedies for privacy violations and data breaches.

At the output stage, GenAI models have the potential to inadvertently expose personal information, thereby raising significant legal and ethical concerns regarding individuals' right to privacy. AI developers commonly rely on biometric data, such as fingerprints, facial features and behavioural patterns, to train and construct biometric recognition models for both public and private sectors.[64] While these models demonstrate high efficiency, they also bring about serious legal apprehensions concerning data breaches[65] and privacy violations.[66] If internet users unknowingly contribute to extensive biometric databases without recognising the associated risks, they may be exposed to substantial privacy infringements risks.[67] The core issue lies in determining the extent of harm suffered by data subjects as a result of such breaches and identifying appropriate remedies.

In recent GenAI lawsuits, privacy violation claims have either been dismissed due to insufficient injury-in-fact standing[68] or settled out of court.[69] Moreover, two unresolved issues

[60] Claudio Novelli and others, 'Generative AI in EU Law: Liability, Privacy, Intellectual Property, and Cybersecurity' (2024) arXiv preprint arXiv:2401.07348, 8.

[61] See, for example, Melissa Heikkilä, 'How to Opt Out of Meta's AI Training' (*MIT Technology Review*, 14 June 2024), <https://www.technologyreview.com/2024/06/14/1093789/how-to-opt-out-of-meta-ai-training/>; Reuters, 'Meta to Start Using Public Posts on Facebook, Instagram in UK To Train AI' (13 September 2024), <https://www.reuters.com/technology/artificial-intelligence/meta-start-using-public-posts-facebook-instagram-uk-train-ai-2024-09-13/> all accessed 18 September 2024.

[62] ibid.

[63] ibid.

[64] Shervin Minaee and others, 'Biometrics Recognition Using Deep Learning: A Survey' (2023) 56(8) Artificial Intelligence Review 8647, 8467–95.

[65] Meng Wang and others, 'Identifying Personal Physiological Data Risks to the Internet of Everything: The Case of Facial Data Breach Risks' (2023) 10(1) Humanities and Social Sciences Communications 1, 1–15.

[66] Matthew B. Kugler, 'From Identification to Identity Theft: Public Perceptions of Biometric Privacy Harms' (2019) 10 UC Irvine L Rev 107, 108–111.

[67] ibid.

[68] *Doe 1 v. GitHub, Inc.*, 672 F.Supp.3d 837, 849–50 (N.D.Cal., 2023).

[69] Kim Lyons, 'Judge Approves $650 Million Facebook Privacy Settlement Over Facial Recognition Feature' *(The Verge*, 27 February 2021), <https://perma.cc/LH4Q-BDSJ> accessed 20 September 2024.

persist in challenging privacy-based claims in GenAI lawsuits: (1) whether the unauthorised use of personal information for training AI models results in concrete injuries or actual harm to privacy rights holders and, if so, (2) how to effectively address and remedy these harms and injuries. Consequently, the growing number of lawsuits targeting GenAI developers not only seek remedies for privacy violations but also attempt to claim restitution for unjust enrichment resulting from the unauthorised use of personal data by developers.

In *P.M. v. OpenAI LP,*[70] a group of plaintiffs initiated a class action lawsuit against OpenAI and Microsoft, alleging that the companies unlawfully scraped a vast amount of private and personal information from publicly available sources and used it without consent to train generative models like ChatGPT and DALL-E.[71] The plaintiffs assert that these actions infringe upon their privacy and property interests in their personal information and data,[72] and they are seeking various remedies, including injunctions, damages and restitution.[73] With regard to the claim of unjust enrichment, the plaintiffs argue that the defendants' profits from using personal information for the commercial training of AI models are unjust and inequitable, and it would be improper to allow such benefits to be derived at the expense of the plaintiffs.[74] To recover their losses, the plaintiffs are seeking restitution for the defendants' unjust enrichment and/or disgorgement of the unjust gains obtained.[75]

In a similar case, *J.L. v. Alphabet Inc*.,[76] the plaintiffs have accused Alphabet's Google of copyright infringement and privacy violations. They claim that Google used personal information and 'copyright materials scraped from websites without authorization,[77] creating an unjust and coercive predicament' for internet users who own the data.[78] The plaintiffs argue that Google's data-scrapping practices lead to unjust enrichment, particularly when AI products like Google's Bard generate revenues from this misappropriated data.[79] The plaintiffs argue that the defendants' misconduct has resulted in 'actual damages' due to the loss of value of their personal information and the lost profits they could have earned from its use.[80] They seek restitution, disgorgement of profits and/or the imposition of a constructive trust to recover the defendants' unjust gains.[81] These lawsuits are ongoing, and any court decision could have significant impact on the conflict between privacy rights holders and AI developers.

In summary, the claim of unjust enrichment could potentially be a cause of action for copyright infringement, violations of the right of publicity and privacy issues, arising from GenAI developers' unauthorised use of data. Data subjects may take advantage of unjust enrichment to pursue more efficient remedies like restitution to recover their losses. However, it remains unclear whether the courts would uphold such claims and provide corresponding remedies. The next part will delve deeper into the determination of unjust enrichment concerning GenAI.

---

[70] *P.M. v. OpenAI LP*, Case No. 3:23-cv-3199 (N.D. Cal. 2023).
[71] ibid.
[72] ibid 64. ('Courts recognize that the internet users have a property interest in their personal information and data. … The economic value of this property interest in personal information is well understood, as a robust market for such data drives the entire technology economy. … Defendants' misappropriation of every piece of data available on the internet, and with it, millions of internet users' personal information without consent, thus represents theft of a value unprecedented in the modern era of technology'.).
[73] ibid.
[74] ibid 143–44.
[75] ibid.
[76] *J.L. v. Alphabet Inc*, Case No. 3:23-cv-03440 (N.D. Cal. 2023).
[77] ibid.
[78] ibid.
[79] ibid 55.
[80] ibid 82.
[81] ibid.

## Unjust enrichment from unauthorised use of data

A claim of unjust enrichment provides a viable alternative for data owners seeking to recover losses incurred from the unauthorised use of their data in GenAI models. This approach requires the consideration of several elements to establish whether unjust enrichment has occurred. Specifically, it must be determined if the defendant was enriched, whether this enrichment was at the claimant's expense, the presence of an unjust factor, the availability of any defences to the claim and the applicable remedies for the claimant.[82]

In cases involving the unauthorised use of protected data by GenAI, the critical inquiries under an unjust enrichment claim generally focus on (1) whether AI developers have indeed gained enrichment from the data, (2) whether this enrichment occurred at the expense of data owners, (3) whether the developers' gain is unjust and (4) whether there are valid defences to the data owners' claims.[83] These considerations form the backbone of assessing claims of unjust enrichment in the context of GenAI, providing a structured framework to address the complexities involved.

### *Enrichment for AI developers*

The first element examines whether AI developers have received benefits through the unauthorised use of protected data. This enrichment can take the form of monetary and non-monetary benefits.[84] Typically, AI developers obtain monetary benefits by reducing the cost associated with data collection. However, the cost of collecting training data has been escalating recently due to the surging popularity of various AI applications. Data brokers and platforms have begun capitalising on this trend by charging premium process.[85] Consequently, smaller AI developers are finding it increasingly challenging to afford the rising licence fee for data usage.[86]

In an effort to mitigate these costs, many developers have sought to avoid paying royalties or licencing fees to data owners. Before the proliferation of AI-related copyright lawsuits, most developers claimed they were using publicly available data from the internet, thereby not infringing on any IP or privacy rights.[87] Although leading AI companies, such as Google and OpenAI have recently started securing high-dollar licencing deals with data brokers or platforms to obtain adequate training data for more powerful AI models, they previously relied on freely scraped data from the Internet to train their AI models. Therefore, data owners contend that it is unjust for AI developers or data brokers to use their data without appropriate compensation. As observed by Posner and Weyl, the value extracted from this data disproportionately benefits a small number of parties, rather than being distributed equitably among the broader populace.[88]

[82] Birks (n 10). The fundamental principle in the US law is similar with subtle deviations. *Jones v. Sparks*, 297 S.W.3d 73, 78 (Ky. App. 2009); see also § 68:5. 'Restitution independent of liability on contract – Unjust enrichment', 26 Williston on Contracts § 68:5 (4th ed.)

[83] Goff and Jones, Birks, and Burrows have each played a pivotal role in clarifying the legal requirements for unjust enrichment claims under English law, developing and promoting a structured and principled framework that has been widely adopted by the courts. Accordingly, this article adopts the four-stage analytical framework as the basis for its legal analysis. To ensure conceptual clarity and consistency, the analysis proceeds with particular reference to Birks' theoretical approach to unjust enrichment.

[84] Birks (n 10) 52–55.

[85] Katie Paul and Anna Tong, 'Inside Big Tech's Underground Race to Buy AI Training Data' *(Reuters,* 6 April 2024), <https://www.reuters.com/technology/inside-big-techs-underground-race-buy-ai-training-data-2024-04-05/>.

[86] Kyle Wiggers, 'AI Training Data Has a Price Tag That Only Big Tech Can Afford'(*TechCrunch*, 1 June 2024) <https://techcrunch.com/2024/06/01/ai-training-data-has-a-price-tag-that-only-big-tech-can-afford/> accessed 20 November 2024.

In cases of unjust enrichment involving non-monetary benefits, the plaintiff normally needs to demonstrate that the defendant is enriched by the receipt of particular good or service[89] In the context of unauthorised use of training data, developers may derive non-monetary benefits by using unauthorised training data to enhance and refine their models, thereby achieving more robust, diverse, and accurate performance capabilities. For example, DALL-E 3 outperforms its predecessor, DALL-E 2, in image creation, partly because its training on more comprehensive datasets that include images owned by human creators.[90] Similarly, the significant advancements in LLMs like Gemini[91] and GPT-4o[92] can largely be attributed to their training on multimodal and multilingual datasets.[93] Further, GenAI models trained on protected data, such as medical records, tend to yield more accurate and reliable results compared to those trained solely on publicly available data.[94]

Overall, AI developers reap both monetary and non-monetary benefits from the enrichment achieved by training their GenAI models on protected data.

### *Enrichment at the expense of data owners*

The second aspect of unjust enrichment pertains to the enrichment obtained at the plaintiff's expense, encompassing any losses, harm or injuries sustained by the plaintiff. In the context of unauthorised data usage to train AI models, it is crucial to establish a connection between the gains acquired by AI developers through the unauthorised use of protected data and the losses, harm and/or injury experienced by the data owners, which could be in economic or non-economic terms.

Data owners, particularly those holding copyright and publicity rights, face substantial losses when they are unable to capitalise on potential revenue from licencing their data to AI developers. Essentially, unauthorised use of copyright works by developers for AI model training deprives owners of the opportunity to generate revenue through licencing agreements. The rise in copyright lawsuits pertaining to AI suggests that copyright holders have incurred financial losses due to their works being used in the training of GenAI models.[95]

[87] See, for example, Saffron Huang and Divya Siddarth, 'Generative AI and The Digital Commons' (2023) arXiv preprint arXiv:2303.11074; Milad Nasr and others, 'Scalable Extraction of Training Data From (Production) Language Models' (2023) arXiv preprint arxiv:2311.17035; Laura Manduchi and others, 'On the Challenges and Opportunities in Generative AI' (2024) arXiv preprint arXiv:2403.00025.

[88] Eric A Posner and E Glen Weyl, *Radical Markets: Uprooting Capitalism and Democracy for a Just Society* (Princeton University Press 2018) 209.

[89] Elise Bant, 'Incapacity, Non Est Factum and Unjust Enrichment' (2009) 33(2) Melbourne U L Rev 368, 374.

[90] OpenAI, 'DALL·E 3' <https://openai.com/index/dall-e-3/> accessed 25 September 2024.

[91] Google, 'Gemini' <https://gemini.google.com/app>.

[92] OpenAI, 'GPT 4o' <https://openai.com/index/hello-gpt-4o/>.

[93] See, for example, Google, 'Prepare Supervised Fine-Tuning Data for Gemini Models' <https://cloud.google.com/vertex-ai/generative-ai/docs/models/gemini-supervised-tuning-about>; Anil Kumar, 'All About Gemini Models and Training Process' (*Medium*, 10 December 2023), <https://theanilbajar.medium.com/all-about-gemini-models-and-training-process-989fc3e25602> all accessed 24 September 2024.

[94] Yan Chen and Pouyan Esmaeilzadeh, 'Generative AI in Medical Practice: In-Depth Exploration of Privacy and Security Challenges' (2024) 26 Journal of Medical Internet Research e53008.

Furthermore, economic loss and non-economic harm may occur when the output of the GenAI models competes directly with protected data, diminishing the latter's market share and the income of data owners.[96] Beyond instances in the copyright market,[97] GenAI models capable of imitating celebrities' identities, such as their likeness and voices, have the potential to deprive them of economic gains and non-economic personal interests generated from their performances.

For instance, in 2025, the United States District Court for the Southern District of New York delivered its judgment in *Lehrman, et al. v. Lovo, Inc.*[98] In this case, two voice actors brought proceedings after discovering that their voices had been appropriated without consent to train an AI algorithm, which was subsequently used to produce synthetic audio products. The court found in favour of the claimants, upholding their successful claim for infringement of the right of publicity.[99]

A similar rationale may be applied to other forms of biometric information generated by AI systems to simulate the identities of celebrities, as these attributes represent core aspects of their persona. Such unauthorised use can detrimentally impact not only the economic income of these celebrities but also their non-economic capital, such as reputation and credibility in the entertainment industries.

Unauthorised data use in GenAI can also infringe upon data owners' non-economic rights and interests, such as moral rights protected under copyright law and personal rights. Studies have shown that incorporating copyright works into AI applications may violate moral rights, including the right of attribution and the right of integrity, thereby causing harm to moral interests of copyright holders.[100]

Regarding personal rights, data owners may suffer reputational harm from the misuse of protected data, such as personal and biometric information, in GenAI models. A prominent case is that of *Clearview AI*, a leading facial recognition technology developer. The company scraped billions of images from social media platforms like Facebook and Instagram to construct an extensive facial database, which was then used to train its facial recognition system for law enforcement purposes.[101] *Clearview AI*'s unauthorised use of biometric information raised

[95] Martin Senftleben, 'Generative AI and Author Remuneration' (2023) 54(10) IIC-International Review of Intellectual Property and Competition Law 1535, 1535–57.

[96] ibid 1556.

[97] See, for example, Jessica L. Gillotte, 'Copyright Infringement in AI-Generated Artworks' (2020) 53 UC Davis L Rev 2655, 2687–90 (discussing the impact of AI-generated works like the New Rembrandt project on the market of original artists' works); Pamela Samuelson, 'Generative AI Meets Copyright' (2023) 381 Science (American Association for the Advancement of Science) 158 (discussing to what extend Stability AI's appropriation of 12 million images from Getty website harms the licencing market).

[98] *Lehrman et al v Lovo, Inc.*, Case 1:24-cv-03770-SPO, (10 July 2025) (SDNY). [*Lovo*]

[99] Ibid.

[100] Rita Matulionyte, 'Can AI Infringe Moral Rights of Authors and Should We Do Anything About It? An Australian Perspective' (2023) 15(1) Law, Innovation and Technology 124, 124–147.

[101] Kashmir Hill, 'The Secretive Company That Might End Privacy as We Know It' (*The New York Times*, 18 January 2020) <https://www.nytimes.com/2020/01/18/technology/clearview-privacy-facial-recognition.html>; Kashmir Hill, 'Clearview AI Used Your Face. Now You May Get a Stake in the Company' (*The New York Times*, 13 June 2024), <https://www.nytimes.com/2024/06/13/business/clearview-ai-facial-recognition-settlement.html> accessed 25 September 2024.

ethical concerns among the public, resulting in privacy violations and reputational damage for individuals whose personal data was exploited.[102] Multiple lawsuits have been filed against Clearview AI in various jurisdictions, including the US,[103] Canada,[104] Australia[105] and the European Union/the UK,[106] alleging breaches of privacy and data protection laws by the company.

In summary, the benefits gained by AI developers through the unauthorised use of protected data lead to both economic losses and non-economic harms for data owners. Therefore, the second element can be substantiated in many of these cases.

### *Unjustness of the enrichment*

This element requires demonstrating that the benefits or enrichment resulting from AI developers' unauthorised use of protected data are unjust or lacks a legal basis.[107] The unjust retention of these gains should be evident in each case.[108] Three commonly recognised unjust factors contributing to unjust enrichment are mistake,[109] failure of consideration[110] and deficiency of consent.[111] The assessment and determination of unjustness can be based on one or more of these factors in specific cases.[112]

Specifically, a mistake as an unjust factor refers to a situation where the claimant caused the enrichment or benefits due to a mistake that occurred in the past or present.[113] The second factor, 'failure of consideration,' arises from the invalidity of the contract under which the parties believed they were operating'.[114] In simpler terms, this means that the defendant received enrichment or benefits without fulfilling their obligations under a valid legal basis. The third factor pertains to the deficiency of consent, meaning that the enrichment or benefits accrued to the defendant without the claimant's consent.[115]

In the context of GenAI, issues of unjustness and mistake often arise when data owners unintentionally allow AI developers to benefit from their data. This typically occurs when data owners, through their online activities, inadvertently grant AI developers access to their data for training future AI models – often by accepting terms of use without fully understanding the implications.[116] This mistake is frequently the result of misunderstanding or a lack of awareness. As previously mentioned, Meta collects data from its user pool via social media platforms and employs this data to train its AI products.[117] Users may not be fully aware of how their data is being utilised to enhance AI models.

---

[102] ibid.
[103] *Calderon v. Clearview AI, Inc.*, No. 20 CIV. 1296 (CM), 2020 WL 2792979 (S.D.N.Y. May 29, 2020).
[104] Clearview AI, Inc., Re, 2021 BCIPC 73, 2021 CarswellBC 3974.
[105] *Clearview AI Inc v. Australian Information Commissioner*, [2023] AATA 1069.
[106] See, for example, European Data Protection Board, 'The French SA Fines Clearview AI EUR 20 Million' (20 October 2022), <https://www.edpb.europa.eu/news/national-news/2022/french-sa-fines-clearview-ai-eur-20-million_en>; Kashmir Hill, 'Clearview AI Successfully Appeals $9 Million Fine in the U.K.' (*The New York Times*, 18 October 2023), <https://www.nytimes.com/2023/10/18/technology/clearview-ai-privacy-fine-britain.html>.
[107] Birks (n 10) 101–102.
[108] George B. Klippert, *Unjust Enrichment* (Butterworths 1983) 139–141.
[109] Birks (n 10) 105
[110] ibid 110.
[111] ibid 105–106.
[112] ibid 106.
[113] ibid 105.
[114] ibid 110.
[115] ibid 116.
[116] Birks (n 10) 110; Klippert (n 108) 176.

Another relevant factor is 'deficiency of consent', which occurs in the collection and exploitation of copyright works for AI training. It is evident that AI developers have benefited from AI products trained on protected data, often without the consent of data owners. This unauthorised use clearly satisfies the unjust factor of 'deficiency of consent'. Although some AI developers have introduced opt-out schemes allowing data owners to exclude their data from future AI model training, they have already utilised a substantial amount of protected data in training earlier models without the owners' permission.[118] Overall, the benefits derived from AI developers' unauthorised use can be deemed unjust when such enrichment meets the unjust factors on a case-by-case basis.

## *Valid defences to data owners' claim*

The fourth element considers whether there are valid defences that can counter claimants' allegations of unjust enrichment.[119] Courts typically evaluate whether these defences could outweigh the unjustness of the defendant's enrichment.[120] While the principle defences for unjust enrichment claims include change of position, bona fide purchase, estoppel, passing on, illegality, statutory limitation and the existence of a contractual or legal justification,[121] this section critically analyses three potential defences, regarding the unauthorised use of protected data, and assesses their validity: contractual or legal justification, statutory limitation and illegality.

The first potential defence involves demonstrating a contractual or legal justification for the use of training data, particularly focusing on whether the data was lawfully obtained from the public domain or through licencing arrangements. AI developers might counter claims of unjust enrichment by demonstrating that their training datasets originate from lawful sources, such as public domain or licenced data. For example, AI developers can legally use data if its owners have made it available through terms of service or Creative Commons licences.[122] However, in the absence of such legal arrangements, it may still be illegal for AI developers to use publicly available data to train their AI models, a common practice in the AI industry.[123] The fact that

[117] Melissa Heikkilä, 'How to Opt Out of Meta's AI Training' *(MIT Technology Review*, 14 June 2024), <https://www.technologyreview.com/2024/06/14/1093789/how-to-opt-out-of-meta-ai-training/>; Reuters, 'Meta to Start Using Public Posts on Facebook, Instagram in UK To Train AI' *(Reuters,* 13 September 2024), <https://www.reuters.com/technology/artificial-intelligence/meta-start-using-public-posts-facebook-instagram-uk-train-ai-2024-09-13/> all accessed 18 September 2024.

[118] Johana Bhuiyan, 'Companies Building AI-Powered Tech Are Using Your Posts. Here's How to Opt Out' (*The Guardian*, 15 November 2024), <https://www.theguardian.com/technology/2024/nov/15/x-ai-gmail-meta-privacy-settings>.

[119] Birks (n 10) 224.

[120] Briks (n 10) 207.

[121] ibid 224–67.

[122] Cade Metz and others, 'How Tech Giants Cut Corners to Harvest Data for A.I.' (*The New York Times*, 6 April 2024), <https://www.nytimes.com/2024/04/06/technology/tech-giants-harvest-data-artificial-intelligence.html> accessed 27 September 2024.

[123] For instance, Google revised its terms of service, particularly its privacy policy, to support the development of its next generation of AI: 'We use publicly available information to help train Google's AI models and build products and features like Google Translate, Bard, and Cloud AI capacities'. See Eli Tan, 'When the Terms of Service

certain data is publicly available does not imply it is free from legal protection and restriction. As mentioned in section 2, unauthorised use of other's data may potentially violate privacy or IP laws.

The second potential defence is based on statutory limitations that preclude restitution. In this context, legal exceptions – such as the fair use and fair dealing in copyright law or exceptions under data protection laws – may offer lawful justification for the unauthorised use of protected data, thereby creating a practical barrier to recovery. For example, the fair use doctrine in the US copyright system allows unauthorised use of copyright works when it satisfies a four-factor test.[124] The proliferation of scholarship has led to controversies regarding whether fair use should extend to cover unauthorised use of copyright works for training AI models.[125]

Proponents who support the adaptability of fair use to GenAI focus on analysing the transformativeness of AI uses by digging into the processes of AI models.[126] Lemley and Casey introduced the concept of 'fair learning', which tailors fair use to machine learning, recognising the necessity of using massive amounts of copyright material.[127] Opponents, however, argue that fair use might be insufficient to address GenAI's copyright infringement issues. Sobel explored two adverse outcomes of applying fair use to AI, which would exacerbate ethical concerns such as discrimination and inequality in society.[128] Li also contended that the 'black box' problem in machine learning impedes the assessment of fair use inquiry.[129]

Judges presiding over pending AI copyright lawsuits in the US have not yet indicated whether fair use would support the practices employed by GenAI. Should they rule affirmatively, fair use could become a robust defence for AI developers against claims of both copyright infringement and unjust enrichment. In the UK, the Copyright, Design, and Patent Act of 2014 includes a TDM exception for non-commercial research.[130] In 2022, the Intellectual Property Office considered expanding this TDM exception to encompass commercial purposes. However, following public consultation, the proposal was withdrawn in 2023.[131] The scope of this exception will undoubtedly influence the viability of defences put forth by AI developers.

The third potential defence is based on the doctrine of illegality, which requires an inquiry into whether the enrichment derived from the unauthorised use of protected data constitutes a benefit arising from unlawful conduct. As previously discussed, unjust enrichment pertains

Change to Make Way for A.I. Training' (*The New York Times*, 26 June 2024) <https://www.nytimes.com/2024/06/26/technology/terms-service-ai-training.html> accessed 27 September 2024.

[124] 17 U.S.C. § 107. The four-factor test considers the purpose and character of the use, the nature of the copyright work, the amount and substantiality of the portion taken and the effect of the use upon the potential market.

[125] See the accompanying text as follow.

[126] Matthew Sag, 'Fairness and Fair Use in Generative AI' (2024) 92 Fordham L Rev 1887; Matthew Sag, 'Copyright Safety for Generative AI' (2023) 61 Hous L Rev 295, 303–313; Michael D. Murray, 'Generative AI Art: Copyright Infringement And Fair Use' (2023) 26 SMU Sci & Tech L Rev 259; Jessica L. Gillotte, 'Copyright Infringement in AI-Generated Artworks' (2020)53 UC Davis L Rev 2655, 2679–2690; Amanda Levendowski, 'How Copyright Law Can Fix Artificial Intelligence's Implicit Bias Problem' (2018) 93 Wash L Rev 579.

[127] Mark A. Lemley and Bryan Casey, 'Fair Learning' (2021) 99 Tex L Rev 743, 748–50.

[128] Benjamin Sobel, 'Artificial Intelligence' Fair Use Crisis' (2017) 41 Colum JL & Arts 45, 79–80, 96–97.

[129] Yangzi Li, 'Does Black-box Machine Learning Shift the US Fair Use Doctrine?' (2021) 16 J Intell Prop L & Prac 1175, 1175–84.

[130] Copyright, Design, and Patent Act (CDPA) 1988, s 29A.

[131] UK Parliament, ' Artificial Intelligence: Intellectual Property Rights vol. 727' (1 February 2023), <https://hansard.parliament.uk/commons/2023-02-01/debates/7CD1D4F9-7805-4CF0-9698-E28ECEFB7177/ArtificialIntelligenceIntellectualPropertyRights> accessed 17 March 2025.

only to benefits attributable to the defendant's wrongful conduct.[132] Beyond the exceptions provided by law, AI developers may counter the claimant's argument by justifying their actions – specifically, the use of protected data without consent – on a legal basis. A notable example is the legality of web scraping to collect training datasets for AI. The lawfulness of web or data scraping varies across jurisdictions and is subject to judicial discretion in specific cases.[133] For example, web scraping is legal under Computer Fraud and Abuse Act (CFAA) in the US.[134] In *hiQ Labs, Inc. v. LinkedIn Corp*, the US Ninth Circuit Court ruled that the defendant's scraping of publicly available information from user profiles did not violate the CFAA, even though the profiles may contain privacy interests.[135] In contrast, EU data protection laws, such as the General Data Protection Regulation (GDPR), restrict the lawful use of web scraping by requiring consent or contractual agreements before utilising others' personal data.[136] Thus, this defence is not valid in every jurisdiction.

In summary, the potential defences outlined in this subsection are not exhaustive. Defences are typically upheld when they can demonstrate that the defendants' enrichment is not unjust or has not occurred at the expense of the claimants.

To conclude section 3, the evaluation of these four elements in determining unjust enrichment within the context of GenAI models is complicated and uncertain. There is no definitive formula to assess whether each element favours the claimant or the defendant in these cases. More importantly, there is significant potential for refining the interpretation and application of the four-element test under the principle of unjust enrichment, allowing more effective analysis of GenAI-related issues. Once the four-element test is established, we must then explore the remedy of restitution for unjust enrichment. The next section will address this topic.

## Unjust enrichment as an alternative solution

The law of unjust enrichment, along with its associated principles of restitution, focuses on recovery of benefits wrongfully obtained, rather than addressing the defendant's wrongful conduct.[137] We argue that unjust enrichment presents a viable solution to the challenges posed by unauthorised data usage in GenAI from two perspectives below.

### *Unauthorised data use as wrongfully obtained benefits*

The principle of unjust enrichment aptly offers a different lens than that provided by IP and privacy laws. As noted above, determining liabilities under these laws necessitates that courts evaluate the lawfulness of AI developers' unauthorised use of data – a process fraught with controversy, particularly regarding the necessity of permission and the scope and limitations of the relevant legal right. Amidst the ongoing debates surrounding the lawfulness of such practices, unjust enrichment can alleviate the negative discourse that may arise from the delegitimisation of unauthorised use, thereby offering some relief to AI innovators.

[132] See accompanying text in Section 2(c).
[133] Andrew Sellars, 'Twenty Years of Web Scraping and the Computer Fraud and Abuse Act', 24 BU J Sci & Tech L 372, 377 (2018) ('Most often the legal status of scraping is characterized as something just shy of unknowable, or a matter entirely left to the whims of courts, plaintiffs, or prosecutors'.).
[134] 18 U.S.C. § 1030(a)(2)(C).
[135] *hiQ Labs, Inc. v. LinkedIn Corp.*, 938 F.3d 985, 995 (9th Cir. 2019).
[136] Ádám Liber, 'The State of Web Scraping in the EU' (3 July 2024), <https://iapp.org/news/a/the-state-of-web-scraping-in-the-eu> accessed 27 September 2024.
[137] Restatement (Third) of Restitution and Unjust Enrichment § 3, 24(2011).

Regardless of the outcome, each judicial decision risks disadvantaging one party. If courts determine that unauthorised use of IP or personal data is wrongful, AI developers may be prohibited from utilising such data without consent, potentially stifling GenAI innovation. Conversely, if courts side with AI developers and permit unauthorised data use, data owners could lose significant opportunities to monetise and maintain control over their data. Such outcomes exacerbate tensions between data owners and AI developers.

In contrast, the law of unjust enrichment adopts a different approach by focusing on the wrongfulness of benefits derived from unauthorised data use. Its primary concern is to assess whether the benefits obtained by the defendant through unauthorised data use are unjust, rather than requiring proof of the wrongfulness of the defendant's actions. This framework not only allows data owners to safeguard their interests without needing to establish the wrongfulness of the developers' action – an issue that remains controversial from both legal and policy perspectives – but also enables AI developers to restore wrongfully obtained benefits to data owners without admitting to any illegality. Thus, this approach can effectively balance the interests of both parties in a manner where copyright and privacy laws may fall short.

### *Restitution as an alternative remedy*

Unjust enrichment applies where other remedies are inadequate or unavailable[138]; therefore, its associated restitution is not part of existing legal, equitable, statutory remedies. For a restitution claim to succeed, the claimant must establish both protected ownership and the defendant's misappropriation.[139] While compensatory damages, probably the most common remedy in private law, are loss based, restitution is a gain-based recovery mechanism.[140] The primary purpose of unjust enrichment is to ensure fairness by addressing the defendant's wrongful gains rather than the plaintiff's losses.[141]

Restitutionary remedies afforded by unjust enrichment more effectively satisfy claimants in certain situations. First, this approach enables the claimant to seek compensation without the burden of proving that the defendant's actions were wrongful and resulted in direct losses or harms. Second, the law of restitution allows recovery of the infringer's profits that exceed the right holder's losses.[142] Third, restitution and unjust enrichment law enable the disgorgement of wrongfully obtained benefits without requiring plaintiffs to prove actual losses or injuries.

IP and privacy laws are the major areas where the unauthorised use of data is debated, and restitution has not been a stranger in these fields.[143] Therefore, restitution can possibly play a role in shaping copyright, privacy and data protection regimes.[144] Under English law, restitutionary and disgorgement remedies for IP violations are rooted in common law and equity.[145] A remedial constructive trust – an equitable remedy established on the grounds of unjust enrichment – may arise in relation to copyright works that have been infringed.[146] Restitutionary disgorgement also serves as a deterrent against 'free riding', where a party benefits from another's IP investments without authorisation. In the realm of privacy, restitution specifically targets economic gains derived by defendants, offering a remedy that is occasionally more effective than compensatory damages.[147] By emphasising economic recovery, restitution incentivises data owners to proactively negotiate their rights, such as copyright and privacy, in a proactive manner.[148]

[138] William Swadling, 'Unjust Enrichment: Value, Rights, and Trusts' (2021)137 LQR56, 56–76.

[139] Restatement (Third) of Restitution and Unjust Enrichment § 42, 33–34 (2011).

[140] Birks (n 10) 11; I. M. Jackman, 'Restitution for Wrongs' (1989) 48 Cambridge LJ 302.

[141] See, for example, Caprice L. Roberts, 'The Case for Restitution and Unjust Enrichment Remedies in Patent Law' (2010) 14 Lewis & Clark L Rev 101, 131; Wendy J. Gordon, 'On Owning Information: Intellectual Property and the Restitutionary Impulse' (1992) 78 Va L Rev 149, 149–282; Wendy J. Gordon, 'Of Harms and Benefits: Torts, Restitution, and Intellectual Property' (2003) 34 McGeorge L Rev 541, 541–70.

[142] Restatement (Third) of Restitution and Unjust Enrichment § 42, 32–33 (2011).

[143] George E. Palmer, *The Law of Restitution* (Little, Brown 1978), §2.9, 120–121; A. Mirshekari, 'Foundations of Legal Protection of Reputation' (2020)11(1) Comparative Law Review 339, 339–361.

[144] See, for example, Roberts (n 141), 131; Gordon (1992) (n 141), 149–282; Gordon (2003) (n 141), 541–70.

The advantages of applying restitution in the context of unauthorised use of protected data to train AI models are obvious. First, because unjust enrichment focuses on the defendant's wrongful gains rather than the plaintiff's injuries, it can alleviate the plaintiff's burden of proof in claiming legal remedies for IP infringement or privacy violations.[149] Second, restitutionary disgorgement, awarded to right holders in cases of IP and privacy violations, ensures recovery whenever the infringer's net profits exceed compensatory damages for the same violation.[150]

While restitution provides a valuable alternative remedy, it is not without challenges. The major challenge is the valuation of intangible assets. Courts have often adopted inconsistent approaches to calculating recoverable profits, particularly in copyright infringement cases.[151] The task of determining restitution based on a defendant's economic gains is fraught with uncertainties, owing largely to the inherently difficult valuation of intangible assets such as data. Similar challenges have arisen in other contexts where the doctrine of unjust enrichment is invoked, prompting several leading scholars to propose practical solutions. For instance, Professor Birk advocated a dual-measure approach for assessing monetary gain, taking into account the 'saving of inevitable expense' and the 'realisation in money of the benefit's otherwise doubtful value'.[152] In a similar vein, Professor Barker introduced a comprehensive six-stage framework for the quantifying gain-based awards.[153]

Moreover, the dynamic nature of AI models significantly exacerbates the difficulty of calculating the benefits accrued by developers from the use of particular datasets. Developers of GenAI models undoubtedly derive value from the unauthorised use of data; however, the diversity and ever-expanding scope of data make it extraordinarily challenging to determine the extent of benefits attributable to specific datasets or data owners. In restitution claims concerning unauthorised use of protected data by AI systems, there is a pressing need for further research – spanning both law and computer science – to develop methodologies capable of adequately addressing these practical challenges.

---

[145] James Edelman, 'Intellectual Property Wrongs' in *Gain-Based Damages: Contract, Tort, Equity and Intellectual Property* (Hart 2002) 217–242.
[146] Lionel Bently and others, Intellectual Property Law (4th edn, Oxford University Press 2014) 1350.
[147] Lauren Henry Scholz, 'Privacy Remedies' (2019) 94 Ind LJ 653.
[148] Restatement (Third) of Restitution and Unjust Enrichment §42, 36–37 (2011).
[149] Chao (n 15) 557.
[150] Restatement (Third) of Restitution and Unjust Enrichment § 42, cmt. g. (2005).
[151] Dane Ciolino, 'Reconsidering Restitution in Copyright' (1999) 48 Emory L. J. 1, 18; Jyh-An Lee, 'Formulating Copyright Damages in China' (2022) 69 J Copyright Soc'y USA185, 188–193.
[152] Birks (n 10) 59–62.
[153] Kit Barker, 'Riddles, Remedies and Restitution: Quantifying Gain in Unjust Enrichment Law' in MA Freeman (ed), *Current Legal Problems*, vol 54 (Oxford University Press 2001) 255, 255–305.

Restitution offers a promising alternative remedy for addressing unauthorised data use in the GenAI context, especially where traditional compensatory damages fall short. By focusing on the defendant's wrongful gains rather than the plaintiff's losses, restitution provides an effective means of recovery and deters future infringements. Restitution has the potential to complement existing remedies, providing a robust framework for protecting data rights in the rapidly evolving AI landscape.

## Conclusion

The unauthorised use of protected data in training GenAI models presents novel challenges to existing legal frameworks, particularly in the areas of IP and privacy law. While these frameworks provide mechanisms to address infringement and privacy violations, they often fall short of adequately balancing the interests of data owners and AI developers. The law of unjust enrichment, which focuses on recovering wrongfully obtained benefits, provides a compelling alternative to address this issue.

Unjust enrichment reframes the debate by shifting focus from the wrongful conduct of AI developers to the benefits they have derived from unauthorised data use. This shift offers several advantages. First, it mitigates the legal uncertainties surrounding the characterisation of unauthorised data use as 'wrongful' under copyright and privacy laws, which often hinges on complex and controversial analyses. Second, by emphasising gain-based recovery, it alleviates the evidentiary burden on data owners, enabling them to seek restitution without needing to establish direct losses or harm. Third, restitutionary remedies have the potential to promote equitable outcomes by ensuring that data owners are compensated for the value of their contributions, while also fostering innovation by allowing AI developers to legitimise their gains through restitution.

### ORCID iD

Jyh-An Lee https://orcid.org/0000-0001-8742-1884

### Funding

The authors disclosed receipt of the following financial support for the research, authorship and/or publication of this article: This work was supported by the Research Grants Council in Hong Kong (grant number CUHK 14608723).

### Declaration of conflicting interests

The authors declared no potential conflicts of interest with respect to the research, authorship and/or publication of this article.